\documentclass[aps,prl,twocolumn,superscriptaddress,showpacs]{revtex4}

\usepackage{graphicx}
\usepackage{dcolumn}
\usepackage{bm}

\def\kpar{\mathbf{k}_\parallel}
\def\SOC{SOC\ }

\begin{document}

\title{Tunneling anisotropic magnetoresistance driven by resonant surface states: First-principles calculations of a Fe(001) surface}

\author{Athanasios N. Chantis}
\thanks{Current address: Theoretical Division, Los Alamos National Laboratory, Los Alamos, New Mexico 87545, USA}
\affiliation{School of Materials, Arizona State University, Tempe, Arizona 85287, USA}
\author{Kirill D. Belashchenko}
\affiliation{Department of Physics and Astronomy and Nebraska Center
for Materials and Nanoscience, University of
Nebraska-Lincoln, Lincoln, Nebraska 68588, USA}
\author{Evgeny Y. Tsymbal}
\affiliation{Department of Physics and Astronomy and Nebraska Center
for Materials and Nanoscience, University of Nebraska-Lincoln,
Lincoln, Nebraska 68588, USA}
\author{Mark van Schilfgaarde}
\affiliation{School of Materials, Arizona State University, Tempe, Arizona 85287, USA}

\date{\today}

\begin{abstract}

Fully-relativistic first-principles calculations of the Fe(001)
surface demonstrate that resonant surface (interface) states may
produce sizeable tunneling anisotropic magnetoresistance in magnetic
tunnel junctions with a single magnetic electrode. The effect is
driven by the spin-orbit coupling.  It shifts the resonant surface
band via the Rashba effect when the magnetization direction changes.
We find that spin-flip scattering at the interface is controlled not
only by the strength of the spin-orbit coupling, but depends
strongly on the intrinsic width of the resonant surface states.

\end{abstract}


\maketitle

Spin-dependent properties of magnetic surfaces and interfaces have
long been a subject of vigorous research. Recent interest is
triggered by the advent of magnetoelectronics, a technology aimed at
harnessing the electrons spin in data storage and processing.
Typical devices utilize heterostructures composed of magnetic and
non-magnetic materials~\cite{zutic}. One approach recently suggested
takes advantage of spin-orbit coupling (SOC), where the resistance
of a nanostructure is controlled by rotating its magnetization with
respect to the interface. For a tunnel junction with only \emph{one}
ferromagnetic electrode this effect is called tunneling anisotropic
magnetoresistance (TAMR)
\cite{gould04,ruster05,giddings05,saito05,giraud05}. It is different
from the usual tunneling magnetoresistance (TMR) observed in
magnetic tunnel junctions \cite{mtj} with \emph{two} ferromagnetic
electrodes when their orientation is switched between the parallel
and antiparallel states. It is also different from the usual
anisotropic magnetoresistance (AMR) typical for bulk ferromagnets
\cite{AMR} and from ballistic anisotropic magnetoresistance (BAMR)
which may occur in ferromagnetic nanocontacts \cite{BAMR}. The TAMR
occurs in the tunneling transport regime because \SOC makes the
electronic structure anisotropic. Practical advantages of TAMR over
AMR are: (1) the tunneling process filters out a fraction of the
electronic phase space, and (2) \SOC is usually stronger at
interfaces.  Both these features result in a larger effect.  An
advantage of a TAMR device over a conventional magnetic tunnel
junction is that only one magnetic interface is necessary.

Experimental demonstrations of TAMR to date have employed diluted
magnetic semiconductors (GaAs:Mn) as magnetic electrodes
\cite{gould04,ruster05,giddings05,saito05,giraud05}.  The TAMR
effect is due to the anisotropy of the tunneling density of states
in the valence band of GaAs induced by \SOC \cite{gould04,ruster05}.
It was suggested that metallic alloys, such as CoPt, with large
magnetocrystalline anisotropy may be used as electrodes in TAMR
devices \cite{shick06}.  TAMR was also observed by scanning
tunneling spectroscopy (STS) of thin Fe films on W(110), as the
result of a group velocity change from the band splitting induced by
\SOC \cite{bode02}.

In this Letter, we propose a different  route to achieve large TAMR
with a metallic electrode.  It is well known that many
transition-metal surfaces, as well as their interfaces with
insulators, exhibit electronic bands that are localized at the
surface or interface.  If such a band does not mix with bulk states
it is called a surface band.  If it mixes weakly with bulk bands, it
broadens and becomes a resonant surface band.  Interface bands
usually contribute strongly to the tunneling current
\cite{wunnicke,stf,coalo,femgo,feomgo}.  On the other hand, when the
magnetization is rotated, these bands shift in the interface
Brillouin zone due to the Rashba effect~\cite{rashba} produced by
\SOC at the interface. For non-magnetic surfaces the Rashba effect
lifts the spin degeneracy of the surface states~\cite{zutic}; for
magnetic surfaces it leads to an asymmetric shift of the
non-degenerate band \cite{krupin05}. The idea pursued in this Letter
is to use this Rashba shift in the TAMR device.  To illustrate this
approach, we choose the Fe(001) surface, because it is one of the
most studied transition-metal surfaces supporting a surface
minority-spin resonant band at the Fermi level \cite{stroscio95}. We
show that the change of the magnetization direction both in and out
of the surface plane results in a sizable change in the tunneling
conductance, which may be observed using STS. We also identify
conditions controlling the strength of spin-flip scattering at the
interface.

To study TAMR produced by electron tunneling from the Fe(001)
surface through vacuum we need a counterelectrode for closing the
electric circuit. For this purpose it is convenient to use a metal
which does not filter the electrons by the transverse wavevector
$\kpar$. A non-magnetic bcc Cu electrode has a spin-independent
free-electron-like band structure and a featureless surface
transmission function \cite{stf}, which makes it an ideal spin
detector, similar to the STS tip in the Tersoff-Hamann theory
\cite{TH}. The semi-infinite Fe and Cu leads are separated by
approximately 1~nm of vacuum (6 monolayers of empty spheres). The
conductance is calculated using the Landauer-B\"uttiker approach
\cite{Datta} implemented  within the fully-relativistic
tight-binding linear muffin-tin orbital (TB-LMTO) method
\cite{chantispre}. Charge self-consistency is achieved using
scalar-relativistic TB-LMTO calculations for Fe and Cu surfaces
treated using supercells with 12 metallic monolayers.

The numerical technique is based on the Green's function
representation of the (TB-LMTO) method in the atomic spheres
approximation \cite{andersen}. Within the relativistic formulation
of the local spin density approximation in which only the spin
component of the current density is taken into account \cite{raja},
inside each atomic sphere we solve the Kohn-Sham Dirac equation
\cite{antrpsov}. The technique is similar to Refs.
\onlinecite{kudr,antrpprb}; the primary difference here is that we
use third-order potential functions \cite{Gunnarson}. The Green's
function of the layered system is constructed by a principal-layer
technique \cite{Turek}, and the conductance is calculated similar to
Ref. \onlinecite{kudr00}. The surface Green's functions of the leads
are constructed scalar-relativistically, and therefore the
conductance $G$ is represented as a sum of four spin components
$G_{\sigma\sigma^\prime}=(e^2/h)T_{\sigma\sigma^\prime}$
\cite{ebert05}, where $T_{\sigma\sigma^\prime}$ is the transmission
function integrated over $\kpar$ (a uniform 200$\times$200 mesh was
used for this integration).

\begin{figure}[tbp]
\includegraphics[width=0.45\textwidth]{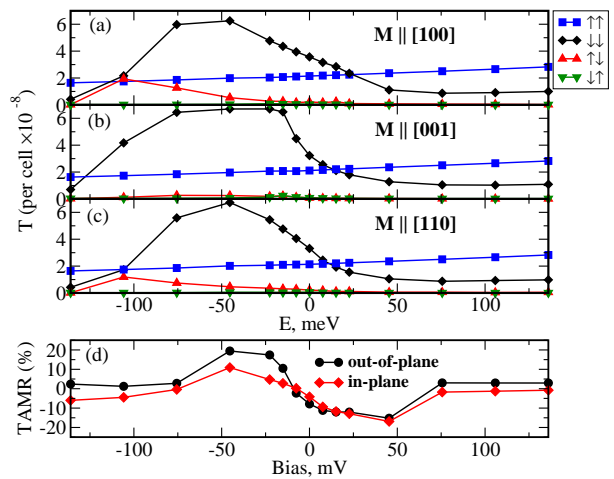}
\caption{ \small (a--c) Spin-resolved integrated transmission
$T_{\sigma\sigma^\prime}$ for the Fe(001) surface with a Cu
counterelectrode as a function of energy. Magnetization is along (a)
[100], (b) [110], and (c) [001] directions. The Fermi level is at
zero energy. (d) In-plane and out-of-plane TAMR as a function of
bias voltage.} \label{TAMR}
\end{figure}

Figs. \ref{TAMR}(a--c) show $T_{\sigma\sigma^\prime}$ for three
magnetization directions. The energy dependence represents the
linear-response conductance in the rigid-band model, which
approximately reflects the effects of alloying. It is seen that
$T_{\uparrow\uparrow}$ exhibits featureless free-electron-like
energy dependence.  However, $T_{\downarrow\downarrow}$ is
nonmonotonic and dominates in the energy range between $-125$~meV
and 25~meV. The TAMR ratios
$(T^{\hat{\bf{n}}}-T^{\bf{[100]}})/T^{\bf [100]}$ are shown in
Fig.~\ref{TAMR}d for both out-of-plane ($\hat{\bf{n}}=[001]$) and
in-plane ($\hat{\bf{n}}=[110]$) magnetization orientations as a
function of bias voltage \cite{bias-note}. In both cases the TAMR
has a spectacular change of sign close to the Fermi level and
reaches $\pm15$--20\% at the bias voltage of $\mp50$~mV.


\begin{figure*}[tbp]
\includegraphics[angle=0,width=0.95\textwidth,clip]{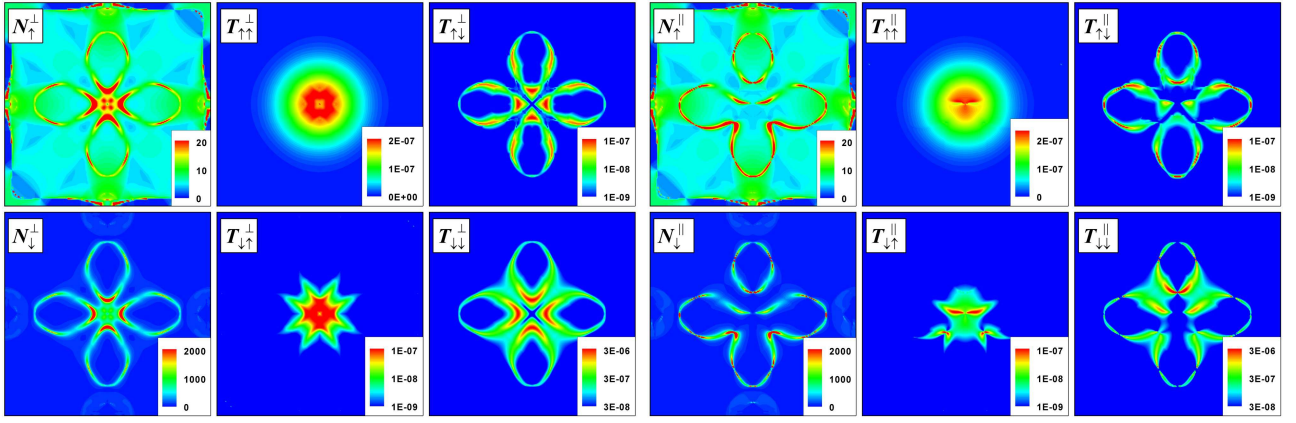}
\caption{Spin components of the DOS $N_\sigma$ at the Fe(001)
surface monolayer and the transmission functions
$T_{\sigma\sigma^\prime}$ for the Fe(001) surface with a Cu
counterelectrode at $15$~meV below $E_F$. Figures are resolved by
$\kpar$ with abscissa along [100] and ordinate along [010]. The left
six panels correspond to the magnetization normal to the surface
(labeled by superscript $\perp$), and the right half to the in-plane
magnetization aligned along [100] (labeled by superscript
$\parallel$). Some panels are given in a logarithmic scale.}
\label{DOS}
\end{figure*}

To explain the origin of large TAMR effect, we focus on energy
$-15$~meV where the conductance has the strongest spin asymmetry.
Fig.~\ref{DOS} shows spin- and $\kpar$-resolved surface densities of
states (DOS) and tunneling transmission functions. The left six
panels correspond to the out-of-plane magnetization
($\hat\mathbf{n}=[001]$), and the right half to the in-plane
magnetization ($\hat\mathbf{n}=[100]$). The resonant surface band is
responsible for the bright four-petal-flower features in the
minority-spin surface DOS ($N^\perp_\downarrow$ and
$N^\parallel_\downarrow$) and in the minority-spin transmission
($T^\perp_{\downarrow\downarrow}$ and
$T^\parallel_{\downarrow\downarrow}$). This band is dominated by the
minority-spin surface states, which mix weakly with bulk bands. A
central point is that \SOC can strongly enhance this mixing, in both
spin-diagonal and spin-mixing components.  In particular, consider
the surface state lying on the $\overline\Gamma\,\overline{X}$ line
with $\kpar=(k_x,0)$. In the absence of \SOC, these eigenstates have
definite parity with respect to reflection in the $y=0$ plane.  The
surface band is even, while the minority-spin bulk band is odd. By
symmetry these states cannot mix, and the surface state remains
localized. The \SOC\ does not conserve this parity and mixes the
surface state with both minority-spin and majority-spin bulk states.
The surface state is thus transformed into a surface resonance. In
our case this occurs at special $\kpar$ points. In general, if for a
given spin the surface band lies within a gap of bulk bands, the SOC
converts this localized surface band into a resonant band.

As is evident from Fig.~\ref{DOS}, the surface bands depend on spin
orientation (compare $N^\perp_\uparrow$ to $N^\parallel_\uparrow$
and $N^\perp_\downarrow$ to $N^\parallel_\downarrow$). This can be
attributed to the Rashba effect \cite{rashba}. The effective
spin-orbit shift of electron energy is given by
\begin{equation}
\Delta \epsilon(\kpar)=\alpha\,(\hat{\bf{z}} \times \kpar) \cdot
\mathbf{s} ,
\end{equation}
where $\hat{\bf{z}}$ is the unit vector normal to the surface.  The
electron spin $\mathbf{s}$ is aligned with the magnetization
$\mathbf{M}$.  For $\mathbf{M}\parallel\hat{\bf{z}}$, the Rashba
shift is zero throughout the surface Brillouin zone, while for
$\mathbf{M}\parallel\hat{\bf{x}}$ it is positive for $k_y>0$ and
negative for $k_y<0$.  The asymmetric shift of the surface bands is
reflected in the loss of fourfold symmetry in the right six panels
of Fig.~\ref{DOS}, and in particular in the loss of mirror symmetry
in $y$.  A similar effect was discussed for the Gd(0001) surface
\cite{krupin05}.

Comparison of $N^\perp_\uparrow$ to $N^\perp_\downarrow$, as well as
$N^\parallel_\uparrow$ to $N^\parallel_\downarrow$, in
Fig.~\ref{DOS} indicates that the admixture of majority-spin states
to the surface band is of the order of 1\% (note the difference in
scales). However, the spin-flip components of the transmission are
quite pronounced.  For example, in certain areas of the Brillouin
zone both $T_{\uparrow\downarrow}$ and $T_{\downarrow\uparrow}$ are
comparable to the majority-spin component $T_{\uparrow\uparrow}$.
For $\mathbf{M}\parallel\hat{\bf{z}}$, a portion of the resonant
surface band lies close to the $\overline\Gamma$ point and adds a
large contribution to the minority-spin conductance.  When the
magnetization is rotated to $\mathbf{M}\parallel\hat{\bf{x}}$ these
states shift, and the conductance is reduced.  This is the origin of
the large TAMR effect seen in Fig.~\ref{TAMR}.

\begin{figure}[tbp]
\includegraphics[width=0.45\textwidth,clip]{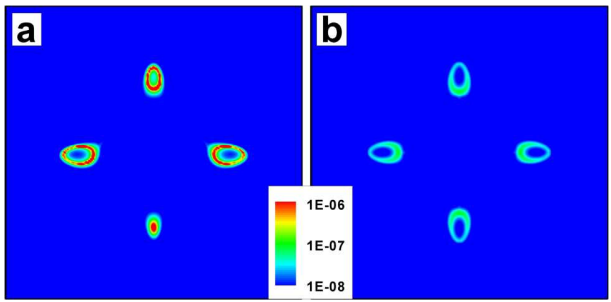}
\caption{ \small Spin-flip transmission $T_{\uparrow\downarrow}$ for
magnetization along (a) [100] (in-plane), and (b) [001]
(out-of-plane), at $-0.102$~meV. } \label{resonant}
\end{figure}

The spin-flip components of the transmission function
$T_{\uparrow\downarrow}$ and $T_{\downarrow\uparrow}$ shown in
Fig.~\ref{TAMR} display a nonmonotonic energy dependence and are
generally quite small compared to the spin-conserving components.
Surprisingly, for the in-plane magnetization,
$T_{\uparrow\downarrow}$ has a pronounced maximum at $E_F-0.1$~eV,
just above the bottom of the resonant surface band, which extends to
higher energies (Fig.~\ref{TAMR}a). Here the spin-flip and
spin-conserving contributions are comparable. Notably, the peak
appears \emph{only} for the in-plane magnetization, and the
spin-flip process is strongly asymmetric: $T_{\uparrow\downarrow}\gg
T_{\downarrow\uparrow}$. Fig.~\ref{resonant} shows $\kpar$-resolved
spin-flip transmission function $T_{\uparrow\downarrow}(\kpar)$ for
both magnetization orientations. The resonant surface bands are seen
as four small ellipses along the $\overline\Gamma\,\overline X$
lines. All the difference in $T_{\uparrow\downarrow}$ for the two
orientations accrues from these four ellipses, which clearly
indicates that the large spin-flip conductance is entirely due to
the resonant surface states.

To elucidate the origin of the resonant spin-flip transmission we
consider a simple tight-binding model. We assume that there is a
single majority band, with surface Green's function
$G^{0}_{\uparrow\uparrow}$ in the absence of spin-orbit coupling,
and a minority surface band $E_s(\kpar)$, broadened by a
hybridization with the bulk minority band. The broadening is
included through parameter $\gamma_0$ that changes the corresponding
Green's function into
$G^{0}_{\downarrow\downarrow}=\left(E-E_s(\kpar)+i\gamma_0\right)^{-1}$.
The spin-orbit interaction $V_{SO}$ mixes spin channels resulting in
the ``dressed'' surface Green's function $G_{\sigma\sigma'}$.

We apply the Landauer-B\"uttiker formalism~\cite{Datta} to the case
of two coupled spin channels, treating the surface monolayer as a
``conductor'' and everything else as ``leads.''  We assume that the
tunneling probability is small and that the second electrode is
non-magnetic.  Then
\begin{equation}
T(\kpar)=\sum_{\sigma\sigma^{\prime}}
T_{\sigma\sigma^{\prime}}=4\sum_{\sigma\sigma^{\prime}}
\mathop{\mathrm{Im}}\Sigma_{\sigma} |G_{\sigma\sigma^{\prime}}|^{2}
\mathop{\mathrm{Im}}\Sigma_{v}, \label{transm}
\end{equation}
where $\Sigma_{\sigma}$ is the spin-dependent self-energy of the
surface layer due to interaction with the magnetic electrode, and
$\Sigma_{v}$ the spin-independent self-energy due to interaction
with the nonmagnetic electrode through vacuum. Including $V_{SO}$
through the Dyson equation, we find the spin-flip component of the
Green's function
\begin{equation}
G_{\uparrow\downarrow}=\frac{V_{SO}G^{0}_{\uparrow\uparrow}}
{E-E_s(\kpar)+i\gamma_0-|V_{SO}|^{2}G^{0}_{\uparrow\uparrow}}.
\label{Gud}
\end{equation}
Using the identity
$\mathop{\mathrm{Im}}G^{0}_{\uparrow\uparrow}${=}$|G^{0}_{\uparrow\uparrow}|^2\mathop{\mathrm{Im}}\Sigma_\uparrow$,
and writing
$-|V_{SO}|^{2}\mathop{\mathrm{Im}}G^{0}_{\uparrow\uparrow}$ as
$\gamma$, we find for the spin-flip transmission coefficient:
\begin{equation}
T_{\uparrow\downarrow}(\kpar)\propto
\frac{\gamma}{\left[E-E_s(\kpar)-\Delta\right]^2+(\gamma_0+\gamma)^2}\,,
\label{TofK}
\end{equation}
where $\Delta=|V_{SO}|^2\mathop{\mathrm{Re}}G^0_{\uparrow\uparrow}$
is the shift of the surface band. Finally, we add an energy
delta-function and integrate over $\kpar$ \cite{note1} to obtain

\begin{equation}
T_{\uparrow\downarrow}(E)\propto
N_s(E)\frac{\gamma}{\gamma_0+\gamma}, \label{Tfinal}
\end{equation}
where $N_s(E)$ is the DOS of the surface band.

The parameters $\gamma$ and $\gamma_0$ can be interpreted as
broadening of the minority surface band due to the coupling to the
majority band (through $V_{SO}$) and to the minority band (through
\emph{both} $V_{SO}$ and hopping matrix elements), respectively.
Eq.~(\ref{Tfinal}) shows that the spin-flip transmission
$T_{\uparrow\downarrow}$ depends on $\gamma/\gamma_0$, and is large
when $\gamma/\gamma_0$ is large.  As was shown above, along the
$\overline{\Gamma}\,\overline{X}$ lines there is no mixing at all
with the minority band when $V_{SO}=0$. Near the bottom of the
interface band, the surface band shrinks to four pockets along these
lines (Fig.~\ref{resonant}), and hence $\gamma_0$ is small for all
the surface states at this energy.

The directional dependence of $T_{\uparrow\downarrow}$
(Fig.~\ref{resonant}) comes from the relative magnitude of the
spin-orbit contributions to $\gamma$ and $\gamma_0$. For the
perpendicular orientation, $\mathbf{M}\parallel\hat{\bf{z}}$,
$V_{SO}$ mixes the interface states primarily to minority-spin bulk
states, contributing mainly to $\gamma_0$.  Therefore
$\gamma\ll\gamma_0$, and the spin-flip transmission is small. For
$\mathbf{M}\parallel\hat{\bf{x}}$, spin-orbit contributions to
$\gamma_0$ and $\gamma$ are of the same order, and resonant
spin-flip transmission sets in. As follows from the Dyson equation,
the second spin-flip component $T_{\downarrow\uparrow}$ is smaller
than $T_{\uparrow\downarrow}$ by a factor of order
$\gamma_0/W_\uparrow$, where $W_\uparrow$ is the bandwidth of the
majority bulk band. Thus, the model predicts
$T_{\downarrow\uparrow}\ll T_{\uparrow\downarrow}$ in perfect
agreement with the detailed calculations, as seen in
Fig.~\ref{TAMR}.

To conclude, we have investigated the electronic structure and
tunneling from the Fe(001) surface to demonstrate that large TAMR
may be achieved by utilizing the Rashba shift of interface resonant
states.  TAMR values of up to $\pm20$\% were predicted for the
Fe(001) surface at small bias voltages; this conclusion may be
checked experimentally using STS with a non-magnetic tip. We also
found a spectacular resonant spin-flip transmission near the
interface band edge, and showed that spin-flip scattering at the
interface depends strongly on the intrinsic broadening of the
resonant band. The results are generally applicable to magnetic
surfaces and interfaces carrying interface states and suggest that
large TAMR may be achieved using metallic electrodes.

\begin{acknowledgments}

The work at UNL was supported by the Nebraska Research Initiative
and NSF MRSEC. The work at ASU was supported by DARPA SPINS project,
and by ONR.

\end{acknowledgments}

\bibliography{fevac}

\begin{thebibliography}{33}
\bibitem[{bia()}]{bias-note}
\bibinfo{note}{The zero-bias conductance is calculated in the window from $E_F$
  to $E_F+$eV. This is an excellent approximation appropriate for comparison
  with STS, because the considered voltages are small.}
\bibitem[{not()}]{note1}
\bibinfo{note}{The integration is first performed in the direction normal to
  the isoenergetic contour, and then along it. Note that the Fermi contour for
  a 2D surface band is the analog of the Fermi surface for a 3D bulk band.}
\end{thebibliography}

\end{document}